\documentclass{article}
\usepackage{fullpage}
\usepackage{color}
\usepackage{verbatim}

\title {On the Interaction of Object-Oriented\\
	Design Patterns and Programming Languages}
\author{Gerald Baumgartner$^*$
        \qquad Konstantin L\"aufer$^{**}$
        \qquad Vincent F. Russo$^{***}$\\[\bigskipamount]
        $^*$ Department of Computer and Information Science\\
	The Ohio State University\\
	395 Dreese Lab., 2015 Neil Ave.\\
        Columbus, OH 43210--1277, USA\\
        gb@cis.ohio-state.edu\\[\bigskipamount]
        $^{**}$ Department of Mathematical and Computer Sciences\\
        Loyola University Chicago\\
	6525 N. Sheridan Rd.\\
        Chicago, IL 60626, USA\\
        laufer@cs.luc.edu\\[\bigskipamount]
	$^{***}$ Lycos, Inc.\\
	400--2 Totten Pond Rd.\\
	Waltham, MA 02154, USA\\
	vrusso@lycos.com}
\date  {February 29, 1996}

\def\del#1{\relax}
\def\new{\textcolor{black}}
\def\note#1{\relax}
\def\draft{
  \def\del{\textcolor{red}}
  \def\new{\textcolor{blue}}
  \def\note##1{\textcolor{green}{[##1]}}
}

\draft

\newcommand{\CPP}{{C\raise.193ex\hbox{\small ++}}}
\newcommand{\GPP}{{G\raise.193ex\hbox{\small ++}}}
\newcommand{\ICPP}{{C\raise.193ex\hbox{\small\em ++}}}
\newcommand{\BCPP}{{C\raise.257ex\hbox{\normalsize\bf ++}}}
\newcommand{\BGPP}{{G\raise.257ex\hbox{\normalsize\bf ++}}}
\newcommand{\LCPP}{{C\raise.193ex\hbox{\LARGE ++}}}

\begin{document}
\maketitle

\begin{abstract}

%\note{
%\begin{center}
%\del{Text to be deleted is typeset in red.}\\
%\new{New text is typeset in blue.}\\
%\note{Comments are in green.}
%\end{center}
%}

Design patterns are distilled from many real systems to catalog common
programming practice. 
However, some object-oriented design patterns are
distorted or overly complicated because of the lack of supporting
programming language constructs or mechanisms. 
For this paper, we have analyzed several published design
patterns looking for idiomatic ways of working around constraints of the
implementation language. 
From this analysis, we lay a groundwork of
general-purpose language constructs and mechanisms that, if provided
by a statically typed, object-oriented language, would better support
the implementation of design patterns and, transitively, benefit the
construction of many real systems.  In particular, our catalog of
language constructs includes subtyping separate from inheritance,
lexically scoped closure objects independent of classes, and
multimethod dispatch.  The proposed constructs and mechanisms are not
radically new, but rather are adopted from a variety of languages and
programming language research and
combined in a new, orthogonal manner.  We argue that by describing
design patterns in terms of the proposed constructs and mechanisms,
pattern descriptions become simpler and, therefore, accessible to a
larger number of language communities.  Constructs and mechanisms
lacking in a particular language can be implemented using {\em
paradigmatic idioms}.

\end{abstract}

\newpage

\section{Introduction}

%\note{In this section, we want to establish the common evolution of
%design patterns and languages.}

For every programming language, there are some (parts of) algorithms
or algorithm structures that cannot be expressed in a straightforward
manner.  For example, using an updatable state or performing system
calls with side effects, such as drawing to the screen or writing to a
file, cannot be expressed conveniently in a purely functional language
such as {\sc Haskell}~\cite{Hudak_et_al._1992}.

In such cases, programmers often develop {\em idioms\/} for expressing
such algorithms or algorithm structures.  In a purely functional
language, for example, any state has to be passed as an argument to
functions accessing the state.  If the state is single-threaded, i.e.,
if only one copy exists of the state, update operations can be
implemented to modify the state destructively without changing the
meaning of the functional program.

%\new{
If a programming idiom is sufficiently common and of general use, a
language construct or mechanism might be introduced into the
programming language to directly support or replace the idiom.  For
example, {\sc Haskell} provides {\em monads\/} to support passing
state to functions at the top level of a functional program~\cite{Peyton_Jones_Wadler_1993}.

%\new{
This evolution of language constructs from idioms can also be observed
in the case of {\em modules}.  To manage the complexity of larger
programs, functions are often grouped with the data on which they work
to form abstract data types.  C~\cite{Kernighan_Richie_1988}, for
example, weakly supports this programming style by distinguishing
between global scope and {\em static\/} file scope.  Auxiliary
functions that should not be accessible to clients of an abstract data
type can be hidden by declaring them to be static.  In languages such
as CLU~\cite{Liskov_et_al._1977} and {\sc Modula-2}~\cite{Wirth_1985},
this program structure has been formalized as the module language
construct.  A definition module is used to declare the module
interface, i.e., it corresponds to global scope in~C\@.  An
implementation module contains the code for the functions declared in
the definition module as well as other, auxiliary, functions.  The
scope of auxiliary functions is limited to the implementation module
and, thus, corresponds to file scope in~C.

Object-oriented programming is now at a stage where enough experience
is available to search for idioms that could be converted into
language constructs or mechanisms.  This programming experience is
made accessible through {\em design patterns}.  Design patterns~\cite{Gamma_et_al._1995,Coplien_Schmidt_1995,Pree_1995} are a
distillation of many real systems for the purpose of cataloging and
categorizing common programming and program design practice.  A design
pattern basically consists of a statement of a design problem together
with the skeleton of a solution.  It is hoped that with the help of a
design pattern catalog, a programmer can reuse proven design solutions
and avoid reinvention.

Since patterns are a reflection of current programming practice, an
analysis of patterns can indicate which language constructs or
mechanisms would be useful in practice and should be provided by new
object-oriented languages.  To understand how an analysis of patterns
gives insight to the language designer, we first need to consider the
influence programming languages have on patterns.

\subsection{Influence of Languages on Patterns}

The choice of language has a two-fold effect on a pattern collection.
First, while many patterns provide genuine design solutions,
as Gamma et al.~observed~\cite[page~4]{Gamma_et_al._1995}, some
low-level patterns can be omitted if the programming language supports
them directly through a language construct or mechanism.
\begin{quote}
\em
The choice of programming language is important because it influences
one's point of view.  Our patterns assume Smalltalk/\CPP-level language
features, and that choice determines what can and cannot be
implemented easily.  If we assumed procedural languages, we might have
included design patterns called ``Inheritance,'' ``Encapsulation,'' and
``Polymorphism.''  Similarly, some of our patterns are supported
directly by the less common object-oriented languages.
\end{quote}
Second, since patterns are distilled from real code, they are often
distorted or overly complicated by idiosyncrasies of the programming
language used, even if the design problem is language independent.

The Singleton pattern~\cite{Gamma_et_al._1995} is an example of a
pattern that merely implements a missing language construct.  A
Singleton is a class with only one instance and code that ensures that
no other instances will be created.  If a language allows the
definition of a single object without a class, or if it offers a
module construct as in Modula-3~\cite{Cardelli_et_al._1992}, there
is no need for the Singleton pattern.

For examples of patterns that work around a defect in the programming
language, consider that most of the complexity of Coplien's
Handle-Body and Envelope-Letter idioms~\cite{Coplien_1992} is due to
the need to implement reference counting since \CPP~\cite{Koenig_1995} lacks garbage collection.  Explicit storage
management also complicates several other idioms in Coplien's
collection.  In a language with a garbage-collected heap, these idioms
would be greatly simplified or not needed at all.

Observe that not all design patterns are influenced by their
implementation language.  Some patterns, such as Producer-Consumer~\cite{Buschmann_Meunier_1995,Meunier_1995}, can be described
independent of the programming language.  The reason the
Producer-Consumer pattern is language-independent is that the
language constructs and mechanism it relies on (arrays or lists and
functions) are basic enough that they can be found in any language.

Many design patterns are not so basic and rely on language mechanisms
such as inheritance, encapsulation, and polymorphism.  This does not
mean that such a pattern is not generally applicable, but rather that
in a non-object-oriented language the missing language constructs and
mechanisms need to be implemented in the form of other patterns.  The
Composite pattern~\cite{Gamma_et_al._1995}, for example, relies on the
ability to use the two classes \verb|Leaf| and \verb|Composite| polymorphically.
To implement subtype polymorphism in C, all that is needed is to
maintain a dispatch table containing function pointers~\cite{Rumbaugh_et_al._1991}.  The Composite pattern can, therefore, be
implemented in C by defining structures representing \verb|Leaf| and
\verb|Composite| together with a function dispatch table implementation of
the Polymorphism pattern.  We term a pattern that implements a missing
language construct or mechanism (such as Polymorphism and Singleton) a
{\em paradigmatic idiom}.

We argue that in order to simplify their presentation and to increase their
audience, object-oriented design
patterns should be described in terms of a {\em richer\/} set of
object-oriented language constructs.
Increasing the number of constructs taken for granted will simplify
many patterns by eliminating distracting implementation details.  Some
existing patterns could even be supported by new language constructs
or mechanisms directly and, thus, become paradigmatic idioms.

To implement a pattern in a
programming language that does not offer the full set of assumed language
constructs, the programmer must combine the pattern with the proper
paradigmatic idioms, as illustrated above.
A catalog of the paradigmatic idioms implementing missing
language features or constructs for an
existing programming language would close the cycle making design patterns
accessible to any language community.

\subsection{Influence of Patterns on Languages}

From the programming languages point of view, patterns are interesting
in that they are a reflection of current programming practice.  An
analysis of patterns, therefore, can indicate which language constructs or
mechanisms would be useful in practice and should be provided by new
object-oriented languages.

Returning to the examples above, the Singleton pattern and its
frequent appearance in other patterns indicates that a module
construct would be a useful addition to object-oriented programming
languages.  Likewise, the complexity of Coplien's idioms to deal with
the lack of an automatically managed heap indicates the need for
garbage collection as a storage management mechanism.

The language constructs and mechanisms we address in this paper are
all related to the object model of a language and its type system.
In particular, we propose that ideal object-oriented languages
\begin{itemize}
\item separate code reuse from subtyping by making inheritance a
      pure code-reuse mechanism and use interface
      conformance to define the subtype relationship,
\item include syntax for specifying lexically scoped closure objects
      independent of the class construct, and
\item provide both single dispatch and multimethod dispatch.
\end{itemize}

By separating the subtyping mechanism from the code reuse mechanism,
we can strengthen the power of both.  In particular, we advocate a
separate language construct for specifying interface types independent
from the construct for defining implementations (the class).  We also
advocate a separate language mechanism for determining the conformance
between classes and interface types (subtyping) independent from the
mechanism for code reuse (inheritance).

In addition to objects that are instances of classes, we propose
lexically scoped, classless closure objects.  Such objects fulfill the
roles that modules, packages, and closures play in other languages.
Furthermore, this notion of objects allows the uniform treatment of a
class's metaclass fields and methods as an object.  By allowing
abstraction over all kinds of objects with explicit interfaces,
polymorphism is obtained in a way that is uniformly applicable to both
singleton (class-less) objects and class instances.

Single dispatch is the appropriate mechanism for adding functionality
to software in the form of new classes while multimethod dispatch is
more appropriate for adding new behavior to a set of existing classes.
By providing both forms of dispatch, software can be evolved in both
of these ways.

Other important mechanisms and constructs, including garbage
collection, synchronization constructs to support concurrency, and
language mechanisms for persistence, distribution, and migration of
objects are not directly related to the object model and type system
of a language and, therefore, are beyond the scope of this paper.

\subsection{Bridging Programming Practice and Language Theory}

We have analyzed design patterns from several sources~\cite{Gamma_et_al._1995,Coplien_Schmidt_1995,Pree_1995,Coplien_1992}
and looked for idiomatic ways of working around constraints of the
implementation language.  Based on this analysis, we catalog
general-purpose language constructs and mechanisms that, if provided
by a statically typed, object-oriented language, would benefit design
patterns and, transitively, a large body of real systems.  We do not
invent radically new language features but rather propose combining
language constructs and mechanisms from a variety of languages in a
new, orthogonal manner.

With this paper, we attempt to bridge programming practice and
language theory.  We base the design of our set of language constructs
and mechanisms on theoretical results but use motivation extracted
through patterns from real-world systems and present it in a language
familiar to practitioners.  Not only does this make the theoretical
results more accessible to practitioners, it is also a confirmation
for theoreticians of the practical usefulness of various results.
Furthermore, we show how to combine these language constructs and
mechanisms by giving an overview of a possible implementation of our object
model.  Our object model together with the overview of its
implementation could, thus, serve as the basis for future language
design.

The following catalog of language constructs and mechanisms is
presented roughly in a pattern style.  Each section starts by
describing a common structuring problem found in patterns, followed by
examples of patterns exhibiting the problem.  We then present a
solution in the form of a language construct or mechanism and conclude
by showing how the solution simplifies or replaces existing patterns.
The paper does not intend to present a complete language design;
rather, code samples and illustrations of proposed language constructs
are presented in pseudo-\CPP\ syntax for illustrative purposes.

\section{Explicit Interface Descriptions}

\label{section:interfaces}

It is often desirable to develop a hierarchy of interface types
independent from the class hierarchy, which provides concrete
implementations of these interfaces~\cite{Canning_et_al._1989,Cook_Hill_Canning_1990}.  With respect to
design, two major problems arise when class inheritance is co-opted
into implementing both the interface and implementation hierarchies
for a system.  First, it becomes difficult to separate logical
abstractions (interfaces) from classes implementing those abstractions
and, second, it becomes difficult or impossible to abstract over
existing code for reuse purposes.

The problem is that most object-oriented languages provide only one
abstraction mechanism: the class.
Implementations of an abstract interface type must explicitly state
their adherence to the interface by inheriting from a class that
declares the abstract interface.  The need to inherit from such {\em
abstract classes\/} often constrains or forces alterations in
implementation hierarchies in order to introduce new interface types.
A separate language construct for abstraction that does not rely on
classes would leave classes free to be used
solely for implementation specification.  If the adherence of a
particular class to an abstract interface type is inferred from the
class specification and does not need to be explicitly coded in the
class, a cleaner separation of interface and implementation can be
achieved. Stating this adherence explicitly might still be useful
for documentation purposes, but should not be required.  A clean,
language-supported separation of interface from implementation also
allows the flexibility of inheritance for code reuse to be
strengthened, as discussed in Section~\ref{sec:inheritance}.

\subsection{Examples}

The need for explicit interface specifications is particularly evident
in the Bridge, Adapter, Proxy, and Decorator patterns~\cite{Gamma_et_al._1995}.

\subsubsection{Bridge and Adapter}

The general idea of the Bridge pattern is to support the construction
of an abstraction hierarchy parallel to an implementation hierarchy
and to avoid a permanent binding between the abstractions and the
implementations of these abstractions.

For example, given the implementation hierarchy,
\begin{quote}
\begin{verbatim}
class Implementor {
public:
    void MethodImp1() = 0;
    void MethodImp2() = 0;
};

class ConcreteImplementorA : public Implementor;
class ConcreteImplementorB : public Implementor;
\end{verbatim}
\end{quote}
we might want to create new classes that delegate parts of their
behavior to classes from the implementation hierarchy:
\begin{quote}
\begin{verbatim}
class Abstraction {
private:
    Implementor * imp;
public:
    void Method1() { imp->MethodImp1(); }
    void Method12() { imp->MethodImp1(); imp->MethodImp2(); }
};

class RefinedAbstraction : public Abstraction;
\end{verbatim}
\end{quote}

The Adapter pattern is essentially the same as the Bridge pattern.
The difference is that the Bridge pattern is used in designs that
separate abstractions and implementations, while in the Adapter
pattern the abstraction is added {\em retroactively}.  We identify the
two patterns equating \verb|Adaptee| with \verb|Implementor| and \verb|Adapter| with
\verb|Abstraction| and choose the name Bridge for both.

The Bridge pattern makes the assumption that all implementation
classes are subclassed from the common abstract superclass
\verb|Implementor| used to define their interface. This can cause problems
in the presence of component libraries that are provided in
``binary-only'' form, or where we desire to use components that are
already part of a library intended for a different application.

The straightforward solution of subclassing the implementation classes
from the abstract superclasses defined by the pattern fails if only
header files and binaries, but no source code, are available for the
two libraries since introducing a new superclass would require access
to the source code of the implementation classes.  The only choices
remaining are to use a discriminated union in the application that
uses the abstractions, to use multiple inheritance to implement a new
set of leaf classes in each implementation hierarchy, or to use a
hierarchy of forwarding classes.  The former solution is rather
inelegant, and the latter two clutter up the name space with a
superfluous set of new class names.  Mularz~\cite{Mularz_1995} makes a
similar observation when discussing building wrappers to access legacy
code.

Even with source code available, if the component classes are already
part of another application, altering their inheritance relationships
could break that application. Again, we are forced to derive new
classes through multiple inheritance or to use forwarding classes.

\subsubsection{Proxy and Decorator}

A more realistic scenario for retroactive abstraction is abstracting
the type of an existing class that is only given in compiled form and
providing an alternate implementation of this type.  If the original
application was not designed with this form of reuse in mind, or if
the alternative implementation uses different data structures, we end
up with a similar problem as above.  The Proxy and Decorator patterns
illustrate this problem.

The purpose of the Proxy pattern is to provide a placeholder for an
object.  For example, an object on a remote machine might be
represented by a proxy on the local machine.  The way this is achieved
in the pattern is by subclassing both the proxy class and the subject
from the same abstract superclass.
\begin{quote}
\begin{verbatim}
class AbstractSubject;
class Subject : public AbstractSubject;

class Proxy : public AbstractSubject {
private:
    Subject * realSubject;
public:
    void Request() { /* ... */ realSubject->Request(); /* ... */ }
};
\end{verbatim}
\end{quote}

The Decorator pattern is identical to the Proxy pattern equating
\verb|Decorator| with \verb|Proxy|, \verb|Component| with \verb|AbstractSubject|, and
\verb|ConcreteComponent| with \verb|Subject|.  The only difference is that while
\verb|Proxy| forwards to a concrete implementation, \verb|Decorator| forwards to
the abstraction \verb|Component|.  The Decorator pattern simply allows us
to ``proxy'' multiple possible implementations.  We identify the two
patterns and use the name Proxy for both.  What distinguishes Proxy
from the Bridge pattern is that the proxy has the same interface as
the object it stands in for (the subject) while the Bridge pattern
defines separate interfaces.

The Proxy pattern would benefit from interfaces separate from classes
since we would not need to introduce \verb|AbstractSubject| as an
abstract superclass. Instead, if the clients of the subject class used
an interface to describe it, a proxy would simply have to implement
that interface and delegate methods through a reference to the
subject. The proxy and the subject would not be constrained to the
same class inheritance hierarchy.  This can be especially important in
cases where the subject class is already in an implementation hierarchy
that might provide fields to the proxy that it did not need or methods
that the client does not ever need to call.

\subsection{Solution}

As we have alluded to, the standard paradigmatic idiom that
implements an interface type hierarchy is a hierarchy of abstract
classes.  Achieving interface hierarchies separate from implementation
hierarchies is problematic in traditional object-oriented programming
languages.  In statically typed languages like \CPP, the required
interface classes are usually defined as abstract base classes~\cite{Ellis_Stroustrup_1990}.  Multiple inheritance (from both an
implementation class and an interface) can be used to establish the
needed type relationships in the implementation hierarchy.

Linking implementation and interface hierarchies together can
lead to type conflicts.  Consider an abstract type \verb|Matrix| with two
subtypes \verb|NegativeDefiniteMatrix| and \verb|OrthogonalMatrix|.  Assume we
wish to have several different implementations of these abstract
interface types, namely \verb|DenseMatrix|, which implements matrices as
two-dimensional arrays, \verb|SparseMatrix|, which uses lists of triples,
and \verb|PermutationMatrix|, which is implemented as a special case
(subclass) of sparse matrices that takes advantage of permutation
matrices having only one element in each row and column.

If we try to model these types and implementations with a single class
hierarchy, we end up either duplicating code or violating the type
hierarchy.  While \verb|DenseMatrix| can be made a subclass of the abstract
classes \verb|Matrix|, \verb|NegativeDefiniteMatrix|, and \verb|OrthogonalMatrix| by
using multiple inheritance, we cannot do the same for \verb|SparseMatrix|.
Since class inheritance normally implies a subtype relationship
between child and parent classes, doing so would make
\verb|PermutationMatrix|, which is a subclass of \verb|SparseMatrix|, an
indirect subclass and, therefore, a subtype, of
\verb|NegativeDefiniteMatrix|.  Since permutation matrices are positive
definite, this would violate the type hierarchy.  The alternative of
having a separate class \verb|SparseNegativeDefiniteMatrix| is not
satisfying either since it causes code replication.

Similar arguments have been given in the literature to show that the
\verb|Collection| class hierarchy of Smalltalk-80~\cite{Goldberg_Robson_1983} is not appropriate as a basis for
subtyping.  While the problem does not arise with dynamic typing, it
becomes an issue when trying to make Smalltalk-80 statically typed
while retaining most of its flexibility.

Explicit support for object interfaces has been introduced into some
object-oriented languages. For example, interfaces are supported in
Java~\cite{Sun_Microsystems_1995} as an explicit syntactical
construct.  A type hierarchy can be created using interfaces, and an
implementation hierarchy can be created through class (single)
inheritance.  However, Java interfaces are in effect only syntactic
sugar for abstract superclasses since implementation classes must
explicitly state which interfaces they implement (akin to inheriting
from an abstract superclass).  Since this type relation is passed down
through the implementation hierarchy, it still can cause the
difficulty described above.

The object form of the Adapter pattern~\cite{Gamma_et_al._1995}
proposes another idiomatic pattern to solve the problem by using a
forwarding class with a single instance variable that references an
implementation class.  This solution is workable in most
object-oriented languages but introduces the added burden of having to
code the forwarding explicitly along with the associated run-time
overhead.

In~\cite{Baumgartner_Russo_1995a,Baumgartner_Russo_1996}, a
conservative extension to \CPP\ is proposed that gives both syntactic
and semantic support for separating interfaces from
implementations. The \verb|signature| construct is much like an abstract
superclass in \CPP, or an interface in Java, except that a class's
conformance to a signature is inferred by the compiler rather than
having to be explicitly declared. This allows much more flexibility in
the implementation hierarchies. Also, since the conformance of a class
to a signature is inferred rather than explicitly declared, it would
be possible to change the semantics of inheritance ({\em not\/} to
imply a subtype relationship) and prevent problems like the one in the
computer algebra example above.  We discuss this further in
Section~\ref{sec:inheritance}.

Thus, the ideal language construct for separating interfaces from
implementations would allow classes and interfaces both to be
specialized through inheritance, and would support inferred subtyping.
In other words, a method invocation on a variable of interface type
should be redirected automatically to an instance of a class
implementing the interface, and the conformance of a class to an
interface should be checked at compile time.

For example, given declarations of the form
\begin{quote}
\begin{verbatim}
interface I {
    void h();
    int g(int);
};

class C {
    // ...
public:
    void h();
    int g(int);
};

class D {
    // ...
public:
    void f(char, int, float);
    int g(int);
    void h();
};

I * ip = new C;
ip->h();
ip = new D;
int j = ip->g(7);
\end{verbatim}
\end{quote}
both class \verb|C| and class \verb|D| conform to the interface \verb|I| and,
therefore, the assignments are valid. Invocations should automatically
be redirected to the proper method implementing the method.

Stating of the conformance to an interface type explicitly might still
be useful for documentation and compile time checking of completeness.
For example, given
\begin{quote}
\begin{verbatim}
class E : implements I {
    // ...
public:
    void h();
    int g(int);
};
\end{verbatim}
\end{quote}
a compiler could check that \verb|E| does indeed implement all the
methods defined in \verb|I|, as it is documented to do so.  However, for
the reasons described above this should not be mandatory.

Another useful feature would be to support method renaming, perhaps
with some form of cast notation.  For example, in the Bridge
pattern~\cite{Gamma_et_al._1995}, one of the libraries may have chosen
different method names.
\begin{quote}
\begin{verbatim}
interface Graphic {
public:
    void draw();
    void move(int, int);
};

class Graphic1 {
public:
    void render();
    void move(int, int);
};

Graphic * g = (rename render to draw) new Graphic1;
\end{verbatim}
\end{quote}
Renaming is useful since it can obviate the need for the Adapter
pattern in cases when the implementation class simply chose a
different name for the method in question than the interface.  When
a simple renaming is not sufficient, class or object Adapters~\cite{Gamma_et_al._1995} can still be used.

\subsection{Uses}

The need for type abstraction separate from implementation is
pervasive throughout numerous patterns (in particular the Structural
Patterns in Gamma et al.~\cite{Gamma_et_al._1995}).  The general
observation can be made that any abstract class that contains no code
should be replaced by an interface definition instead. For example, in
the Abstract Factory pattern, the \verb|AbstractProduct| classes could all
be interfaces as would be the \verb|AbstractFactory| class itself; in the
Builder pattern, the \verb|Builder| class would be an interface; in the
Bridge pattern, the \verb|Implementor| class would be an interface; in the
Decorator pattern, the \verb|Component| class should be an interface; etc.
Using interfaces instead of classes also better documents the uses of
the abstractions.
		% Vince
\section{Improved Code-Reuse Mechanisms}

\label{sec:inheritance}

The construction of a class can frequently be simplified by reusing
code from existing classes.  Inheritance is the characteristic reuse
mechanism provided by object-oriented languages.  In fact, the
possibility for code reuse offered by inheritance is one of the
reasons for the popularity of object-oriented languages.

Code reuse by inheritance can be grouped into four different patterns:
specializing an existing class, filling in missing pieces in a
framework, composing a class from existing classes, and creating a new
class from {\em pieces\/} of existing classes.%
\footnote{To our knowledge, the only form of code reuse that has been
published as a pattern is Template Method.  The other forms of code
reuse seem to be considered directly supported by a language's
inheritance mechanism.  Since inheritance is assumed to be a basic
language mechanism~\cite{Gamma_et_al._1995}, these code reuse patterns
are not found in object-oriented design pattern collections.  We
suggest to include all four code reuse patterns in pattern collections
as a guide for programmers to use inheritance properly.}

\paragraph{Specialization}
For specializing an existing class, a subclass can add new fields and
methods or override existing methods.  In \CPP\ terminology, the
mechanisms used for this purpose are {\em public inheritance\/} and
redefinition of {\em virtual\/} member functions~\cite{Koenig_1995}.
In Smalltalk-80, all inheritance is public and all methods are virtual~\cite{Goldberg_Robson_1983}.

\paragraph{Template Method}
A {\em framework\/} is a skeleton of an application or algorithm that
implements the control structure of a class of related applications or
algorithms.  The framework defines an interface for pieces to be
filled in to create a specific instance of such an application or
algorithm.  Aside from overriding virtual methods, a mechanism
commonly used in frameworks is an {\em abstract\/} method.  An
abstract method is a method that can be called by other methods of a
framework but is not implemented in the framework class itself.  An
implementation has to be provided by a subclass.  The Smalltalk-80 term
for this mechanism is {\em subclass responsibility}.  The style of
refining certain steps of an algorithm is described in the Template
Method pattern in~\cite{Gamma_et_al._1995}.

\paragraph{Mixin}
In the mixin programming style~\cite{Moon_1986}, inheritance is used
to add functional components to a class.  The same effect could be
achieved by making the components fields of the class.  However,
making the methods of a component available to clients of the
class would require writing forwarding methods.  Mixin
inheritance simplifies the reuse of the components whose public
methods should be made available.  The standard mechanisms used for
this style of code reuse is {\em multiple inheritance\/} of (mostly)
{\em non-virtual\/} methods.  The import statement in Modula-3~\cite{Cardelli_et_al._1992} serves
a similar purpose, except that imported functions are not
automatically re-exported.

\paragraph{Theft}
In some cases, a class might only inherit part of its superclass(es)
or rename inherited methods purely for reusing existing code without
intending any semantic relationship between the superclass(es) and the
subclass.  In \CPP, the mechanism used for this purpose is {\em
private\/} inheritance.  In Modula-3, it is possible to import
functions from a module selectively.

\vspace{\baselineskip}

Since inheritance in most object-oriented languages also defines a
subtype relationship, the possibilities for code reuse have to be
restricted such that the subtype relationship is not broken.
In particular, the definition of public inheritance in \CPP\ is
slightly limited, which limits the applicability of the Specialization
pattern.  Since private inheritance does not define a subtype
relationship, the Theft pattern is often avoided.

\subsection{Examples}

Since we have interface conformance for defining a subtype relationship,
we can strengthen inheritance for code-reuse purposes by breaking the
subtype relationship it usually defines.  This makes the
Specialization pattern more flexible, as the following example
demonstrates.  Not having to use inheritance for defining a subtype
relationship also allows programmers to use the Theft pattern more
often.

To motivate the limitation of public inheritance, suppose we have an
implementation of coordinate points and would like to extend them to
color points (this example is based on examples from~\cite{Cook_Palsberg_1989,Cook_Hill_Canning_1990}).
Class \verb|ColorPoint| inherits class \verb|Point|, adds color support, and
redefines equality to compare the color as well.
\begin{quote}
\begin{verbatim}
class Point {
protected:
    int x, y;
public:
    virtual double dist() { return sqrt (x * x + y * y); }
    virtual int equal(Point & p) { return (x == p.x) && (y == p.y); }
    virtual Point & closer(Point & p) { return (p.dist() < dist()) ? p : *this; }
    // ...
};

class ColorPoint : public Point {
protected:
    Color c;
public:
    virtual int equal(ColorPoint & p) { return (x == p.x) && (y == p.y) && (c == p.c); }
    // ...
};

ColorPoint & p = // ...
ColorPoint & q = // ...
int i = p.equal(q.closer(p));           // type error
\end{verbatim}
\end{quote}
The last line results in a compile-time type error, since \verb|equal()|
expects its argument to be of type \verb|ColorPoint&| but the return type
of \verb|closer()| is of type \verb|Point&|.  By comparing whether \verb|q| or \verb|p| is
closer to the origin, we have lost the information that both are color
points.

What we would like is a form of inheritance that results in
\verb|ColorPoint::closer()| having the type
\begin{quote}
\begin{verbatim}
virtual ColorPoint & closer(ColorPoint &);
\end{verbatim}
\end{quote}
That is, we would like occurrences of type \verb|Point| in an parameter type or
return type to be changed to \verb|ColorPoint| when a method is inherited.
\CPP\ does not allow this since changing parameter types in this
fashion would violate the contravariance rule required for subtyping.

\subsection{Solution}

If subtyping is achieved by testing the conformance of
a class to an interface, we do not need inheritance for
subtyping purposes.  Assume class \verb|D| is a subclass of \verb|C| defined by
\CPP-style public inheritance and overriding of virtual methods.
Since \verb|D| conforms to any interface that \verb|C| conforms to, an instance
of \verb|D| can be assigned to an interface reference wherever an instance
of \verb|C| was assigned before.  The run-time dispatch for method calls
through an interface reference gives the same polymorphism as a method
dispatch.  That is, interface conformance subsumes any subtype
relationship defined by inheritance in which only virtual methods are
overridden.  The only case where interface conformance cannot be used
instead of class inheritance to define a subtype relationship is when
the subclass overrides a {\em non-virtual\/} method of the superclass.
This use of inheritance, however, is usually considered a programming
error~\cite{Cline_Lomow_1995}.

If inheritance does not define a subtype relationship, it can be made
more flexible for code reuse purposes in two ways.  By introducing the
notion of \verb|selftype|, it is possible to override methods covariantly.
Furthermore, private inheritance can be used more frequently for code
reuse.

The type \verb|selftype| refers to the receiver's class type.  If a
declaration involving \verb|selftype| is inherited, \verb|selftype| is rebound
to refer to the subclass.  For example, in
\begin{quote}
\begin{verbatim}
class Point {
protected:
    int x, y;
public:
    virtual double dist() { return sqrt(x * x + y * y); }
    virtual int equal(Point & p) { return (x == p.x) && (y == p.y); }
    virtual selftype & closer(selftype & p) { return (p.dist() < dist()) ? p : *this; }
    // ...
};
\end{verbatim}
\end{quote}
\verb|selftype&| is synonymous with \verb|Point&|.  In class \verb|ColorPoint|,
inheriting class \verb|Point| results in the parameter and return types
of \verb|closer| to be synonymous with \verb|ColorPoint&|.  The call to \verb|equal| in
\begin{quote}
\begin{verbatim}
ColorPoint & p = // ...
ColorPoint & q = // ...
int i = p.equal(q.closer(p));
\end{verbatim}
\end{quote}
now type-checks correctly.  In interface inheritance, \verb|selftype|
refers to the interface type.

Using \verb|selftype| together with virtual methods gives the programmer
more flexibility in specializing an existing class at the expense of
losing the subtype relationship.  For some applications, however, this
is the desired behavior.  If a subtype relationship is needed, it can
be achieved by declaring parameter and return types to be of an
interface instead of a class type.

For implementing frameworks using the Template Method pattern~\cite{Gamma_et_al._1995}, a subtype relationship between the
abstract class and the concrete class is not required.  The concrete
class inherits the code of the abstract class and fills in the missing
methods.  The result is that the concrete class conforms to any
interface the abstract class conforms to.

Since both for specializing an existing class and for defining a
framework virtual methods are more common, we suggest to use syntax
similar to Java's~\cite{Sun_Microsystems_1995}: methods are virtual
by default unless they are explicitly declared \verb|final|.  Also for
specifying abstract methods, a keyword, such as Java's \verb|abstract|
keyword, could be used.

To support the mixin programming style, we suggest that
object-oriented languages support an import mechanism
related to that found in Modula-3.  Importing a class works
similar to inheritance, except that all methods of the imported class
are treated as non-virtual methods, whether they were declared final
or not.  In other words, importing a class \verb|C| is operationally the
same as defining a field of type \verb|C| and writing forwarding methods
for all of \verb|C|'s public and protected methods.

To support unstructured code reuse, i.e., the Theft pattern, we can
allow renaming of inherited methods or only inheriting part of a
superclass.  For example, with a syntax such as
\begin{quote}
\begin{verbatim}
class CDE :
    private C only f1, f2;
    public D except f;
    public E only f, g rename g to E_g;
{
public:
    // ...
};
\end{verbatim}
\end{quote}
we could inherit the methods \verb|C::f1| and \verb|C::f2| without re-exporting
them, inherit everything of class \verb|D| except the method \verb|D::f|, and
inherit \verb|f| and \verb|g| from class \verb|E| while renaming \verb|E::g| to \verb|E\_g| to
avoid a name conflict with \verb|D::g|.  The same syntax can be used for
selectively importing a class.  Similar syntax has been proposed in~\cite{Mitchell_Meldal_Madhav_1991}.

Note that in the above example, the interface of class \verb|CDE| will
conform to the interface of class \verb|D| if \verb|D::f| and \verb|E::f| have the
same type.  Interface conformance, therefore, enables us to define
subtype relationships that could not be achieved with inheritance.

If classes \verb|D| and \verb|E| in the example above both inherit from class
\verb|A|, an instance of class \verb|CDE| would have two copies of the fields of
class \verb|A|.  For classes \verb|D| and \verb|E| to share the same \verb|A| part, in
\CPP\ they would need to use {\em virtual inheritance\/} from class \verb|A|.
We suggest to use {\em sharing constrains\/} similar to those in ML~\cite{Milner_Tofte_Harper_1990,Milner_Tofte_1991} instead.  In the
above example, we could specify the sharing constraint
\begin{quote}
\begin{verbatim}
    sharing D::A == E::A;
\end{verbatim}
\end{quote}
in class \verb|CDE|.  The advantage over virtual inheritance is that its
use does not have to be anticipated when defining classes \verb|D| and \verb|E|.

\subsection{Uses}

In most patterns in Gamma et al.~\cite{Gamma_et_al._1995},
inheritance is used to define a subtype relationship by inheriting an
abstract superclass.  In these cases, the proper language construct to
use would be an interface.

The most common form of code reuse found in design patterns is a
framework.  The Template Method pattern explains how to build a
framework.  Further frameworks are found in Factory Method, Singleton,
Adapter, Mediator, and Observer.  The \verb|Component| and \verb|Decorator|
classes in the Composite and Decorator patterns also represent
frameworks.

The only cases in Gamma et al.~of using inheritance to specialize an
existing class are in the Decorator and Mediator patterns.  The
refined abstractions in the Bridge pattern would be implemented as
specializations of an interface.

Only the class adaptor in the Adaptor pattern uses private inheritance
as a short-hand for composition.  The object adaptor is structurally
much cleaner.

In all cases where inheritance is used both for type abstraction and
for code reuse, using interfaces for the abstractions simplifies the
inheritance structure and makes it unnecessary for inheritance to
define a subtype relationship.
		% Gerald
\section{Objects Without Classes}

\label{sec:modules}

Many applications contain only a single copy of certain components.
For example, a compiler contains only one parser.  Even though there
might be multiple network interfaces, an operating system contains
only one TCP/IP stack.  In the Abstract Factory pattern~\cite{Gamma_et_al._1995}, there is only one product factory.

In languages that only provide classes for constructing objects, it is
necessary to ensure that certain classes get instantiated at most
once.  This is the purpose of the Singleton pattern~\cite{Gamma_et_al._1995}.

In addition to constructing singleton software components, there is a
need for packaging components for program delivery.  Some
object-oriented languages provide constructs specifically for this
purpose.  Examples are namespaces in \CPP~\cite{Koenig_1995} or
packages in Java~\cite{Sun_Microsystems_1995}.  Shaw~\cite{Shaw_1995}
lists module as a pattern for component packaging.

Namespaces in \CPP\ and packages in Java provide rudimentary support
for packaging but are not flexible enough to be used as singleton
components.  They cannot be parameterized, passed to functions, or
specialized through inheritance.

\subsection{Examples}

\subsubsection{The Singleton Pattern}

Instead of defining a singleton object, using the Singleton pattern~\cite{Gamma_et_al._1995} allows the programmer to define a class that
has only one instance.  To ensure that no more than one instance can
be created, it is necessary to intercept requests to create new
objects.  Using a named class allows the singleton object to be passed
to functions or to be specialized in subclasses.

In \CPP, a typical implementation of the Singleton pattern is (from~\cite{Gamma_et_al._1995}):
\begin{quote}
\begin{verbatim}
class Singleton {
public:
    static Singleton * Instance();
protected:
    Singleton();
private:
    static Singleton * _instance;
};
\end{verbatim}
\end{quote}
By not making the constructor public, the only way for clients to
create an instance is through the method \verb|Instance()|:
\begin{quote}
\begin{verbatim}
Singleton * Singleton::_instance = 0;

Singleton * Singleton::Instance() {
    if (_instance == 0) {
        _instance = new Singleton;
    }
    return _instance;
}
\end{verbatim}
\end{quote}
Making \verb|Instance()| a class method and storing the single instance in
a class field ensures that only one instance can be created.
Accessing the object indirectly through the \verb|Instance()| method has
the disadvantage that it precludes static resolution of method calls
to the singleton object.

\subsubsection{Component Packaging}

The Abstract Factory pattern~\cite{Gamma_et_al._1995} provides an
interface for creating families of products.  Suppose we need to
create scrollbars and windows of either the Presentation Manager or
the Motif product families.  A skeleton of an implementation for these
products might look as follows:
\begin{quote}
\begin{verbatim}
interface AbstractScrollBar;
class PMScrollBar : implements AbstractScrollBar;
class MotifScrollBar : implements AbstractScrollBar;

interface AbstractWindow;
class PMWindow : implements AbstractWindow;
class MotifWindow : implements AbstractWindow;
\end{verbatim}
\end{quote}
A concrete factory, would create products from one of the two
families, e.g., \verb|MotifScrollBar|s and {\tt Motif\-Window}s.
\verb|AbstractScrollBar| and \verb|AbstractWindow| define common interfaces for
both product families.

If an application has to deal with many different product families,
the naming conventions for the different types of products quickly
become unmanageable.  What is needed is a packaging mechanism that
introduces a new scope and allows grouping, e.g., \verb|ScrollBar| and
\verb|Window| into a \verb|Motif| package.

\subsection{Solution}

% While \CPP\ allows to declare a singleton object as an instance of
% a nameless class,
% \begin{quote}
% \begin{verbatim}
% class {
% public:
%     // ...
% } global;
% \end{verbatim}
% \end{quote}
% such objects are not very useful.  Since they don't have a
% named type, they cannot be passed to functions (except by supressing
% type checking using the ellipsis).  They also cannot be specialized by
% inheritance.
%
% These shortcomings of nameless classes can be overcome with the
% Singleton pattern at the expense of explicitly managing the creation
% of the object.

The bookkeeping effort of explicitly managing the creation of
singleton objects can be avoided if the language provides syntax for
constructing objects without requiring a class to be declared.  The
same syntax would allow packaging components.

\subsubsection{Singleton Objects}

A simple language design solution to support the singleton pattern
would be to introduce an {\em object construct\/} with the same syntax
as a class construct but without constructors or destructor.  In
pseudo-\CPP\ syntax, the declaration
\begin{quote}
\begin{verbatim}
object Singleton1 {
public:
    // public methods
private:
    // private data
};
\end{verbatim}
\end{quote}
could be used to define and initialize the constant \verb|Singleton1|.

With such syntax, it is guaranteed that only one object will be
created and, therefore, references to it can be statically resolved.
In addition, unlike namespaces in \CPP\ or packages in Java, these
objects can be extended by inheritance.
\begin{quote}
\begin{verbatim}
object Singleton2 : public Singleton1 {
public:
    // additional public methods
private:
    // additional private data
};
\end{verbatim}
\end{quote}

To abstract over singleton objects, we can use the interface construct
described earlier.  It is not necessary to provide separate interface
constructs for objects and classes.  Given an interface type \verb|T|, a
reference of type \verb|T&| can then be assigned either a singleton object
that provides all the methods specified in \verb|T| or an instance of a
conforming class:
\begin{quote}
\begin{verbatim}
interface T {
    // ...
};

T & p = Singleton1;
T * q = new C;
\end{verbatim}
\end{quote}

Interface types also allow us to define polymorphic functions that
can take as argument any singleton conforming to the interface.
\begin{quote}
\begin{verbatim}
int f(T &);
int i = f(Singleton1);
int j = f(Singleton2);
\end{verbatim}
\end{quote}
If clients require access to only one of several subobjects of
\verb|Singleton1|, all objects could be encapsulated in an outer object that
exports a reference to only one of the subobjects.

\subsubsection{Packages}

The object construct described above is related to {\em module\/}
constructs as found in Modula-3~\cite{Cardelli_et_al._1992} or ML~\cite{Milner_Tofte_Harper_1990,Milner_Tofte_1991}.  Unlike modules in
these languages, objects are first-class values, i.e., they can be
passed to and returned from functions or assigned to variables.  The
advantage of modules is that they allow packaging and exporting of
types in addition to variables and functions.  Modules can therefore
be considered {\em higher-order\/} objects.

Both Modula-3 and ML provide different constructs for module
interfaces (called \verb|INTERFACE| in Modula-3 and \verb|signature| in ML)
and for the implementation of a module (called \verb|MODULE|
in Modula-3 and \verb|structure| in ML).  The components exported from the
module are those listed in the interface.

In \CPP-style syntax, the interface and implementation parts could be
combined into one construct and labeled with \verb|public| and \verb|private|,
respectively.  Syntactically, a package then looks the same as a
singleton object:
\begin{quote}
\begin{verbatim}
object Motif {
public:
    class ScrollBar : implements AbstractScrollBar { /* ... */ };
    class Window : implements AbstractWindow { /* ... */ };
    // other public types, data, and methods
private:
    // private types, data, and methods
};
\end{verbatim}
\end{quote}

The main difference is that packages also export types, i.e., packages
are higher-order objects.  (We use the terms ``package'' and
``singleton object'' to distinguish between the two uses of the
\verb|object| construct.)  A package interface can be defined using the
same interface construct as for classes.

In addition to modules, ML provides parameterized modules called
\verb|functor|s.  In \CPP-like syntax, a parameterized package could be
defined as a template:
\begin{quote}
\begin{verbatim}
template <T1 ComponentClass, T2 Package1> object Package2 : implements T3 {
    // ...
};

Package2<C, P> pkg;     // C must conform to T1, P must conform to T2
\end{verbatim}
\end{quote}
where \verb|T1| and \verb|T2| are interface types constraining the possible
argument classes and packages, respectively, and \verb|T3| is the interface
of the resulting package.  For type-checking template parameters and
results, package interfaces can also specify the types that a package
must export.  Such higher-order interfaces are only useful for type
checking templates.

To allow information hiding, both Modula-3 and ML allow exported types
to be {\em opaque}.  An opaque type is a type whose name is declared
in the module interface but whose definition is only given in a module
implementation.  Clients of a module, which only get access to the
name of an opaque type, can declare variables of the type, initialize
them by calling a function of the module, and pass them to other
functions.  Clients cannot inspect or alter values of an opaque type,
except through functions exported by the module.

The style of programming with modules and opaque types is known as the
{\em abstract data type (ADT) style}.  The role of opaque types in
this programming style corresponds to private fields in an
object-oriented style.  In this pseudo-\CPP\ syntax, an opaque type
would be a type whose {\em name\/} is declared \verb|public| but whose
definition is \verb|private|.

Sometimes it is useful for packages to be passed to a function or
method by reference (for an example, see
Section~\ref{sec:metaclasses}).  However, exporting types complicates
the assignment to an object reference.  Given a package interface \verb|T|,
suppose the function `\verb|int f(T& P)|' takes as argument a package that
exports the non-opaque type \verb|t|.  Inside the function, it would then
be possible to declare a variable of type \verb|P.t| and, since \verb|P.t| is
not opaque, it would be possible to inspect a value of this type.
However, the exact type may not be known at compile time since it can
depend on the actual package being passed.  Allowing a function to
inspect values of a non-opaque type would, therefore, require types to
be first-class values and some type checking to be done at run-time.

The solution for keeping the language statically typed has
traditionally been to make modules second class, i.e., to disallow
module references.  Neither ML nor Modula-3 allows passing modules to
functions.

The solution for getting the best of both worlds, exported types {\em
and\/} first-class packages, is to consider exported types opaque when
a package is accessed through an object reference.  Such a solution
was proposed as an object-based extension of ML modules by Mitchell et
al.~\cite{Mitchell_Meldal_Madhav_1991} and for supporting separate
compilation of ML modules by Harper and Lillibridge~\cite{Harper_Lillibridge_1994}.

As with classes, it is often useful to define new objects by reusing
the code of existing objects, as shown in the definition of
\verb|Singleton2| above.  Since all of the code reuse patterns in
Section~\ref{sec:inheritance} apply to objects as well, all forms of
inheritance and import can be used for defining objects.  The only
exception is \verb|selftype|, since singleton objects and packages do not
have an implementation type.  An object containing an \verb|abstract|
method would need to be considered an abstract object and could only
be used as a parent for object inheritance.

In summary, to benefit packaging and pattern implementations, we
suggest adding to object-oriented languages an ML-style module system
consisting of (parameterized) packages and package interfaces.  As
with classes and class interfaces, the conformance of a package to a
package interface could be tested either structurally or by name.  A
package that does not export types, i.e., a singleton object, is a
first-class value and can be assigned to references or passed to
functions.  A package that does export types is a higher-order object.
When passing such as package to functions, all exported types become
opaque to make the package first-class.  Like classes, packages or
singleton objects can be refined by inheritance.

The proposed language construct would replace the Singleton pattern
and provide superior facilities for packaging components as compared
to \CPP's namespaces or Java's packages.

\subsection{Uses}

In a language with singleton objects as described above, the Singleton
pattern is no longer needed.  An extended Singleton pattern that
allows a variable number of instances can be modeled by controlling
the instantiation of parameterized objects.  For languages without
classless objects, the Singleton pattern would be the necessary
paradigmatic idiom.

Any pattern that uses the Singleton pattern can be expressed directly
using the \verb|object| construct.  In the Abstract Factory pattern~\cite{Gamma_et_al._1995}, the abstract factory and the abstract
product become interfaces, and the concrete factory becomes an object.
Only the concrete products remain classes.  Similarly in the Builder
pattern~\cite{Gamma_et_al._1995}, builder and concrete builder become
an interface and an object, respectively.

The purpose of the Facade pattern~\cite{Gamma_et_al._1995} is to
package software components.  Since usually only one facade object is
required, it could be implemented as a package.  Similarly, a Mediator~\cite{Gamma_et_al._1995} would typically be a package.

Both the Strategy and the Visitor patterns~\cite{Gamma_et_al._1995}
package only methods.  They do not define any new data structure.
Concrete strategies and concrete visitors would therefore be singleton
objects, with the abstract classes \verb|Strategy| and \verb|Visitor| being
replaced by interfaces.  Similarly, each concrete state in the State
pattern~\cite{Gamma_et_al._1995} would be a singleton object.
			% Gerald
\section{Lexically Scoped Closure Objects}

\label{section:closures}

Some programming situations call for a {\em (lexical) closure\/}
mechanism for creating behavior on the fly that can be invoked at a
later time but has access to the lexical environment current when this
behavior was created.  Common situations that could benefit from a
closure mechanism are the parameterization of an object by behavior,
the state change of an object from one behavior to another, and the
creation of new behavior by partial application of existing behavior.

Many statically typed object-oriented languages such as \CPP~\cite{Ellis_Stroustrup_1990} or Java~\cite{Sun_Microsystems_1995}
neither allow behavior to be created on the fly, nor do they give
functions or objects access to the surrounding local environment.  By
contrast, Smalltalk-80~\cite{Goldberg_Robson_1983} provides {\em
blocks} as a limited form of closures.

%Edwards~\cite{Edwards_1995} explains how to implement streams with objects.

%Iterators are bogus.  Better use closures, lambdas, and multimethods.

%The argument that the interface of iterators should be kept
%separate from the collection (list) doesn't count, since they are
%so tightly coupled, that it amounts to only one interface anyway.
%The collection needs to create the iterator and the iterator needs
%to access private data of the collection (or the collection needs
%a separate interface to access the data).

%It's difficult to make multiple iterators polymorphic.  Adding
%an iterator requires changing the factory method in the collection.

%To make iterators robust to collection changes, the collection
%needs special support to tell the iterator.

%registration -- register in constructor, unregister in destructor
%argument for automatic but explicit finalization

%It's difficult, if not impossible, to build multiple, polymorphic,
%and robust iterators.

%The only advantage of external iterators is that more than one can
%operate at the same time.

%Internal iterators need closures/lambdas.

%Better solution: use mapping functions (`internal iterators')
%exported by the collection or combine collection and mapping
%functions in an enclosing module.  Lazy streams could be used
%instead of external iterators.  We could build an `external
%iterator' from a `internal iterator' (mapping function) plus
%the functions delay and force.

\subsection{Examples}

Specialized behavioral patterns such as Command, Iterator, State, and
Strategy~\cite{Gamma_et_al._1995} are workarounds for missing language
support for lexical closures.  The abundance of such patterns has
probably emerged due to a lack of closure support in many
object-oriented languages and a failure to identify closures as the
underlying generalized mechanism.  We first illustrate the need for
lexical closures by examining the Iterator pattern.  We then argue
that the remaining three patterns can be unified under the concept of
behavior encapsulated in a closure object that has access to the
lexical environment current when the object was created.

\subsubsection{The Iterator Pattern}

The purpose of the Iterator pattern is to support efficient iteration
over a collection without exposing its internal structure.  This
pattern allows a collection to be traversed in different ways without
requiring the collection to provide a method for each kind of
traversal.  Furthermore, the Iterator pattern supports multiple
simultaneous traversals over the same collection.

The Iterator pattern actually describes an {\em external\/} iterator,
i.e., the control flow for the traversal is outside of the iterator or
container operations.  By contrast, an {\em internal\/} iterator is
usually provided as a method of the container class that takes a
visitation function as its argument.  For example, the \verb|do:| method
in the Smalltalk-80 collection class hierarchy~\cite{Goldberg_Robson_1983} is an internal iterator parameterized by a
block.

The following example shows a typical application of the Iterator
pattern for multiple traversal of a container in \CPP.  Each of the
two iterators keeps track of one position in the traversal.  The
\verb|ListIterator| class provides operations for initializing the
traversal, accessing the current item, going to the next element, and
checking whether more items follow.  The \verb|ListIterator| class is a
friend of the \verb|List| class to allow direct access to the
representation of the \verb|List| class.
\begin{quote}
\begin{verbatim}
template <class Item> class List {
public:
    // ...
    friend class ListIterator<Item>;
};

template <class Item> class ListIterator {
public:
    void First();
    void Next();
    bool IsDone();
    Item & CurrentItem();
    // ...
};

List<int> aList;
// ...
ListIterator<int> i(aList), j(aList);
for (i.First(); ! i.IsDone(); i.Next()) {
    for (j.First(); ! j.IsDone(); j.Next())
        cout << setw(8) << i.CurrentItem() * j.CurrentItem();
    cout << endl;
}
\end{verbatim}
\end{quote}

%template <class Item> interface Iterator {
%public:
%    void First();
%    void Next();
%    bool IsDone();
%    Item CurrentItem();
%};

This pattern has a number of weaknesses.  First, encapsulation is
sacrificed for efficiency reasons, since the iterator class is a
friend of the collection class.  Second, the user is required to write
his or her own control structure instead of using a method that
already encapsulates the control flow appropriate for the traversal.
This means that the user must strictly follow a protocol for the order
in which the iterator methods are invoked although this protocol
cannot be expressed in the interface of the iterator.  For example,
invoking \verb|Next| after \verb|IsDone| returns true is probably undefined.
Finally, robustness becomes an issue with external iterators because
the user can change the container by adding or deleting elements
during iteration~\cite{Kofler_1993}.

% Kof 93] Kofler, T.: Robust Iterators for ET++. Structured Programming,
% Vol. 14, No. 2, 1993, pp. 62-85.

Internal iterators do not have these problems, but are hard to write
in languages like \CPP\ because of the lack of lexical scoping.
External iterators are mostly a workaround for the lack of internal
iterators in such languages~\cite{Baker_1993}.  The only other
advantage of external iterators is that they allow the user to control
the progress of the iteration.  This is useful in situations in which
it is not desirable to traverse an entire collection at once (see also
the pairwise iteration example in Section~\ref{section:multimethods}).

\subsubsection{The State, Strategy, and Command Patterns}

The purpose of the State, Strategy, and Command patterns is to
encapsulate behavior in an object.

Specifically, the State and Strategy patterns support configuring
objects with one of several possible behaviors.  Both patterns consist
of a client class called \verb|Context| and a behavioral class called
\verb|State| or \verb|Strategy|, respectively.  The only difference between
\verb|State| and \verb|Strategy| is that \verb|State| allows dynamic configuration,
while \verb|Strategy| usually provides static configuration.  There is also
a minor stylistic difference in the \CPP\ code given for the two
patterns~\cite{Gamma_et_al._1995}.  In the \verb|State| pattern, the
\verb|Context| class does not use the \verb|State| class in its public
interface.  In the \verb|Strategy| pattern, by contrast, the constructor of
the \verb|Context| class is explicitly parameterized by a \verb|Strategy| class
or object.  Given the behavioral and structural similarities, we
identify the two patterns and choose the name Strategy for both.

The Command pattern applies the Strategy pattern to encapsulate a
request with its receiver.  The pattern consists of a client class
called \verb|Invoker| and a behavioral class called \verb|Command|.  

In all three patterns, the behavioral classes have subclasses
\verb|ConcreteState|, \verb|ConcreteStrategy|, and \verb|ConcreteCommand|,
respectively.

Gamma et al.~do not establish any relationship between the three
patterns other than relating State and Strategy to the Flyweight
pattern~\cite{Gamma_et_al._1995}.  
%Zimmer notices the similarities and, in trying to analyze them, 
%reinvents Strategy and calls it Objectifier~\cite{Zimmer_1995}.  
Zimmer recognizes the commonalities between these behavioral patterns
and tries to capture them in the Objectifier pattern~\cite{Zimmer_1995}, which is not significantly different from the
Strategy pattern.

%Neither Gamma et al.~nor Zimmer
%mention that instances of the Strategy pattern are often singletons.

The following example (adopted from~\cite{Gamma_et_al._1995})
illustrates the Strategy pattern.  A composition maintains a
collection of textual and graphical components of a document.  Upon
creation, a composition can be parameterized with the desired layout
strategy to be invoked in the \verb|Layout()| method.
\begin{quote}
\begin{verbatim}
class Composition {
public:
    Composition(Compositor *);
    void Layout();
private:
    Compositor * _compositor;
    Component * _components;
};

void Composition::Layout() {
    CompositionData theCompData;
    // ...
    _compositor->Compose(theCompData);
    // ...
}
\end{verbatim}
\end{quote}
The \verb|Compositor| interface describes layout strategies that determine
how the components of a document should be arranged into lines.  A
\verb|SimpleCompositor| looks at components one line at a time to decide
where line breaks should go.  An \verb|ArrayCompositor| breaks the
components into lines each containing a fixed number of components.
\begin{quote}
\begin{verbatim}
interface Compositor {
    int Compose(CompositionData & data);
    // ...
};

class SimpleCompositor : implements Compositor { /* ... */ };
class ArrayCompositor : implements Compositor { /* ... */ };
\end{verbatim}
\end{quote}
Finally, when a new document is created, the desired layout strategy
is passed as an argument.
\begin{quote}
\begin{verbatim}
Composition * quick = new Composition(new SimpleCompositor);
Composition * table = new Composition(new ArrayCompositor);
\end{verbatim}
\end{quote}

%class SimpleCompositor : implements Compositor {
%public:
%    int Compose(CompositionData & data);
%    // ...
%};
%
%class ArrayCompositor : implements Compositor {
%public:
%    int Compose(CompositionData & data);
%    // ...
%};
%
In this example, implementations of the \verb|Compositor| interface are used as
behavioral arguments, that is, as a replacement for closures, which
are not directly supported by languages such as \CPP.

\subsection{Solution}

The problems addressed by the patterns discussed in this section and
other specialized behavioral patterns are naturally solved by
introducing a closure mechanism for creating new behavior that
captures the environment current at the time this behavior was created.
Such behavior comes either in the form of a single function or an
entire object bundling several methods.  Functions or objects can then
be defined in any scope and use identifiers defined in the current
environment, which consists of definitions in outer scopes.  The key
idea is that the function or object might be passed around and used
elsewhere, but retains access to portions of the current environment.
This kind of mechanism is known as {\em (lexical) closure\/} and is a
standard feature of functional or applicative languages and of some
object-oriented languages, including Smalltalk-80~\cite{Goldberg_Robson_1983}, CLOS~\cite{Bobrow_et_al._1988, Steele_1990, Paepcke_1993}, and Cecil~\cite{Chambers_1992, Chambers_1993}.

Coplien~\cite{Coplien_1992} and L\"{a}ufer~\cite{Laufer_1995}
developed a paradigmatic idiom that simulates closures in \CPP\ using
classes.  Their approaches fall short of true lexical closures in that
closures cannot be anonymous and must capture explicitly the portions
of the environment they use.  Breuel~\cite{Breuel_1988} describes an
extension of \CPP\ that supports efficient named lexically scoped
functions.  We propose further generalizing those ideas by introducing
lexically scoped closure objects.

We use the object construct previously introduced in
Section~\ref{sec:modules} to express closure objects.  A closure
object can be named or anonymous.  A closure object can be created on
the fly and captures the surrounding lexical environment current at
the time the object was created.  A closure object is allowed to
escape the lexical scope in which it was created and takes the
captured environment with it.  As in most functional languages,
supporting closures requires allocating on or moving to the heap the
activation records of any functions or methods that allow a closure to
escape their scope.

The following example includes an interface \verb|Counter| for simple
counter objects and a function {\tt Make\-Counter} that creates counter
objects with a given initial setting.  When a counter object is
created, it captures the variable \verb|value| from the environment current
at that time.
\begin{quote}
\begin{verbatim}
interface Counter {
    void Next();
    int Current();
};

Counter & MakeCounter(int value) {
    return object {
        void Next() { value++; }
        int Current() { return value; }
    };
}

Counter & from100 = MakeCounter(100);
from100.Count();
cout << from100.Current();      // prints "101"
\end{verbatim}
\end{quote}

We observe that a function closure is merely a closure object with a
single function call method (e.g., \verb|operator()| in \CPP).  It might be
useful to allow function notation as {\em syntactic sugar} for this
common case.  The next example shows a counter function that increases
the counter each time the function is invoked.
\begin{quote}
\begin{verbatim}
typedef int (* Counter) ();

Counter MakeCounter(int value) {
    return int (*) () { return value++; };
}

Counter from100 = MakeCounter(100);
cout << from100();              // prints "100"
cout << from100();              // prints "101"
\end{verbatim}
\end{quote}

%whole thing becomes more compositional

The following is a version of the multiple traversal example that uses
internal iterators and closures instead of external iterators.  It is
usually hard to perform nested iterations over the same collection
using internal iterators.  Using lexical closures, such iterations
become easy and natural.  The key observation is that we can pass to
the inner iteration a closure that has access to the state of the
outer iteration.
\begin{quote}
\begin{verbatim}
template <class Item> interface Action {
    void Apply(Item i);
};

template <class Item> class List {
public:
    void ForEach(Action<Item> & a);
    // ...
};

List<int> aList;

aList.ForEach(
    object {
        void Apply(int i) { 
            aList.ForEach(
                object {
                    void Apply(int j) { cout << setw(8) << i * j; }
                }
            )
            cout << endl;
        }
    }
)
\end{verbatim}
\end{quote}

Using function notation, the previous example can be rewritten as
%Lexically scoped functions could be considered merely special cases
%of objects with a single function call method (e.g., \verb|operator()| in
%\CPP).  Function notation could be allowed as {\em syntactic sugar}, as
%shown in the following version of the previous example.
%%
%\begin{quote}
%\begin{verbatim}
%template <class Item> class List {
%public:
%    typedef void (* Action)(Item);
%    void ForEach(Action a);
%    // ...
%};
%
%List<int> aList;
%
\begin{quote}
\begin{verbatim}
aList.ForEach(
    void (*) (int i) { 
        aList.ForEach(void (*) (int j) { cout << setw(8) << i * j; })
        cout << endl;
    }
)
\end{verbatim}
\end{quote}

Similar to the iterator example above, many applications of the
Strategy pattern and its variations can be expressed more naturally by
creating a function that invokes a method from an object defined in
the current environment.  The function itself can be invoked from
elsewhere.
\begin{quote}
\begin{verbatim}
MyClass * receiver = new MyClass;
// ...
void aCommand() { receiver->MyAction(); }
// ...
otherFunction(aCommand);
\end{verbatim}
\end{quote}

\subsection{Uses}

Workarounds for the lack of lexical closures in a language are as
pervasive~\cite{Coplien_1992, Laufer_1995} as attempts to introduce
them into languages that lack them~\cite{Breuel_1988, Dami_1994}.
Closures and closure objects are useful for any type of behavioral
parameterization, including callbacks in event-driven systems and
parameters to applicative versions of iterators as used in functional
languages~\cite{Kuehne_1995}.

%%% Local Variables: 
%%% mode: latex
%%% TeX-master: "Paper"
%%% End: 

		% Konstantin
\section{Metaclass Objects}

\label{sec:metaclasses}

Many patterns rely on the ability either to abstract over classes or
to parameterize an operation based on an object's class.  For example,
it would be useful to allow object construction to be parameterized by
a class chosen at run time.  It is also often desirable to abstract
over methods that operate on the class itself rather than on an
instance of the class.  An example of such a method is a method
accessing a class instantiation count.

To achieve abstraction over classes, we need an object representing
the class at run time.  Such {\em metaclass objects\/} would have
fields and methods associated with the class, such as instantiation
counts or constructors.  We could then abstract over metaclass objects
with regular interfaces, treat classes as values, and even allow
parameterization by a class.

\subsection{Examples}

\subsubsection{Abstract Factory and Factory Method}

The Abstract Factory pattern~\cite{Gamma_et_al._1995} provides an
interface for creating families of products without specifying their
concrete classes.  Suppose we want to configure an application with
either the Presentation Manager or the Motif look-and-feel.  When
creating widgets such as scrollbars or windows, the application should
not hard-code the names of the widget classes.

Given the following interfaces and classes,
\begin{quote}
\begin{verbatim}
interface AbstractScrollBar;
class PMScrollBar : implements AbstractScrollBar;
class MotifScrollBar : implements AbstractScrollBar;

interface AbstractWindow;
class PMWindow : implements AbstractWindow;
class MotifWindow : implements AbstractWindow;
\end{verbatim}
\end{quote}
we need to separate from the application the code for instantiating
the classes.  The application only knows that the generated objects
conform to the interfaces \verb|AbstractScrollBar| and \verb|AbstractWindow|,
respectively.

The solution suggested by the Abstract Factory pattern is to introduce
an interface for factories creating the widgets.
\begin{quote}
\begin{verbatim}
interface WidgetFactory {
    AbstractScrollBar & CreateScrollBar ();
    AbstractWindow & CreateWindow ();
};
\end{verbatim}
\end{quote}
In \CPP, \verb|WidgetFactory| would be defined as an abstract class.  For
each product family, a concrete factory class is needed to generate
the proper widgets:
\begin{quote}
\begin{verbatim}
class PMWidgetFactory : implements WidgetFactory {
public:
    AbstractScrollBar & CreateScrollBar () { return new PMScrollBar; }
    AbstractWindow & CreateWindow () { return new PMWindow; }
};

class MotifWidgetFactory : implements WidgetFactory {
public:
    AbstractScrollBar & CreateScrollBar () { return new MotifScrollBar; }
    AbstractWindow & CreateWindow () { return new MotifWindow; }
};
\end{verbatim}
\end{quote}
At startup time, one of the concrete factories is selected.  The
application refers to the factory through a reference of type
|\verb|WidgetFactory|.

The Factory Method pattern~\cite{Gamma_et_al._1995} is similar to the
Abstract Factory pattern but refers to only one creation method.  In
the above example, both \verb|CreateScrollBar()| and \verb|CreateWindow()| are
factory methods.  The structure of the Factory Method pattern is often
to make the application a framework class, called \verb|Creator|, with the
factory method as an abstract method that needs to be supplied by a
subclass \verb|ConcreteCreator|.

The disadvantage of both patterns is that the different product
families need to be mirrored by different concrete factories and
concrete creators, respectively.  With a run-time representation of
the product classes, we could parameterize both the abstract factory
and the creator, with the product class(es) to be instantiated.

\subsection{Solution}

Smalltalk-80~\cite{Goldberg_Robson_1983} supports classes as objects
directly and, as discussed in Gamma et al.~\cite{Gamma_et_al._1995},
such metaclasses greatly simplify cases like the one described above.
\CPP\ only supports a limited form of metaclass by allowing fields and
variables to be declared \verb|static|.  For example,
\begin{quote}
\begin{verbatim}
class A {
private:
    static int i;
public:
    static int geti() { return i; }
    A() { /* ... */ }
    // ...
};
\end{verbatim}
\end{quote}
declares that all instances of \verb|A| share a common field \verb|i| and the
method \verb|geti()|.  Such metaclass data is not encapsulated in an object
but rather accessed using the scope resolution syntax.  In this
example, \verb|i| and \verb|geti()| are accessed as \verb|C::i| and \verb|C::geti()|,
respectively.

The problem with static members as an approximation of metaclasses is
that \CPP\ does not allow the use of classes as values. If the static
parts were encapsulated in an object, we could assign metaclass
objects to references, invoke metaclass methods through these
references and, most importantly, abstract over metaclass objects
using interfaces.

Ideally, any syntax for declaring metaclasses should attempt to group
the methods and fields of the metaclass object and the methods and
fields of instances of the class into one construct in order to
localize class documentation and simplify maintenance.  A simple
\CPP-like syntax for merging metaclass definitions and instance
definitions into one construct might be
\begin{quote}
\begin{verbatim}
class A {
meta private:
    int i;

meta public:
    A & new();
    A & new(int i, String s);
    int geti() { return i; }

public:
    int f();
    void g(int);
};
\end{verbatim}
\end{quote}
This construct would simultaneously define both the metaclass object
and the implementation type of class instances.

Note that the object constructor becomes the method \verb|new()| of the
generated metaclass object.  Making the constructor a method of the
metaclass object with a well-known name allows us to abstract over
object creation, as needed for the Abstract Factory and Template
Method patterns, and to statically type-check this abstraction.

Abstracting over metaclass objects can be accomplished with interfaces
that name the metaclass methods but not the instance methods.  This
abstraction is unambiguous.  For example, while
\begin{quote}
\begin{verbatim}
interface instanceA {
    int f();
    void g(int);
};
\end{verbatim}
\end{quote}
abstracts over the instances of class \verb|A| above,
\begin{quote}
\begin{verbatim}
interface metaA {
    instanceA & new();
    instanceA & new(int, String);
};
\end{verbatim}
\end{quote}
abstracts over its metaclass object. Note that in this interface the
return type of \verb|new| is \verb|instanceA| and not \verb|A|. This is necessary to
enforce the encapsulation provided by class \verb|A| and to allow the
|\verb|metaA| interface type to abstract over other implementations of the
same interface as well.

Consider again the Abstract Factory pattern.  Since
the \verb|ConcreteFactory| classes do nothing but create instances of
products from the proper product families, it would be possible to
replace them completely with an \verb|AbstractFactory| class parameterized
by the product family's classes. With a statically type-checked
metaclass object (as with the object interfaces discussed in
Section~\ref{sec:modules}), it is also possible to check that only the
correct \verb|Product| classes are created by methods of \verb|AbstractFactory|.

For example, we can define interfaces for the products,
\begin{quote}
\begin{verbatim}
interface AbstractScrollBar {
    void handleClick();
};

interface AbstractWindow {
    void move(int x, int y);
};
\end{verbatim}
\end{quote}
and then define interfaces for classes that generate objects of types
|\verb|AbstractScrollBar| and {\tt Abstract\-Window}, respectively:
\begin{quote}
\begin{verbatim}
interface ScrollBarMaker {
    AbstractScrollBar & new(int x, int y, String label);
};

interface WindowMaker {
    AbstractWindow & new(int x, int y, int w, int h);
}
\end{verbatim}
\end{quote}
Implementations of the product classes can either implicitly or
explicitly implement the product and metaclass interfaces, as in
\begin{quote}
\begin{verbatim}
class PMScrollBar {
meta public:
    PMScrollBar & new(int x, int y, String label) { /* ... */ }
public:
    void handleClick() { /* ... */ }
};

class PMWindow : implements AbstractWindow {
meta public:
    PMWindow & new(int x, int y, int w, int h) { /* ... */ }
public:
    void move(int x, int y) { /* ... */ }
};

class MotifScrollBar : implements AbstractScrollBar {
meta public:
    MotifScrollBar & new(int x, int y, String label) { /* ... */ }
public:
    void handleClick() { /* ... */ }
};

class MotifWindow {
meta public:
    MotifWindow & new(int x, int y, int w, int h) { /* ... */ }
public:
    void move(int x, int y) { /* ... */ }
};
\end{verbatim}
\end{quote}
A factory is then just a singleton object that is initialized with
the product classes to instantiate. For example,
\begin{quote}
\begin{verbatim}
object WidgetFactory {
private:
    ScrollBarMaker & scrollbarMaker;
    WindowMaker & windowMaker;

public:
    void init(ScrollBarMaker & scrollbars, WindowMaker & windows) {
        scrollbarMaker = scrollbars;
        windowMaker = windows;
    }

    AbstractScrollBar & makeScrollBar(int x, int y, String label) {
        return scrollbarMaker.new(x, y, label);
    }
    AbstractWindow & makeWindow(int x, int y, int h, int w) {
        return windowMaker.new(x, y, h, w);
    }
};

WidgetFactory.init(MotifScrollBar, MotifWindow);
AbstractWindow & win = WidgetFactory.makeWindow(0, 0, 100, 100);
\end{verbatim}
\end{quote}
In this example, the singleton object \verb|WidgetFactory| fulfills the
roles of both abstract and concrete factories in the pattern.  The
static type-checking of metaclass interfaces guarantees that only the
proper types of classes will be used as the products. For example, a
window class could not be passed in as the scrollbar class to
instantiate in \verb|makeScrollBar| since a window class would not match
the interface required by \verb|ScrollBarMaker|.

Using objects to package implementations, we can simplify this pattern
even further and allow whole component libraries to be checked.  For
example, we can package together all related product implementations:
\begin{quote}
\begin{verbatim}
object Motif {
public:
    class ScrollBar {
    meta public:
        ScrollBar & new(int x, int y, String label) { /* ... */ }
    public:
        void handleClick() { /* ... */ }
    };

    class Window {
    meta public:
        Window & new(int x, int y, int w, int h) { /* ... */ }
    public:
        void move(int x, int y) { /* ... */ }
    };
};
\end{verbatim}
\end{quote}
and then provide an interface description of such a package:
\begin{quote}
\begin{verbatim}
interface ProductLibrary {
    AbstractScrollBar & ScrollBar.new(int x, int y, String label);
    AbstractWindow & Window.new(int x, int y, int w, int h);
};
\end{verbatim}
\end{quote}
using syntax that allows to specify methods of nested classes.

This avoids many problems.  \verb|ProductLibrary| establishes an
interface for all component libraries to be used with the factory.  It
prevents, for example, Motif scrollbars from being used together with
Presentation Manager windows.  The factory simply becomes a singleton
object parameterized by the product library from which to instantiate.
\begin{quote}
\begin{verbatim}
object WidgetFactory {
private:
    ProductLibrary & P;

public:
    void init(ProductLibrary & lib) {
        P = lib;
    }

    AbstractScrollBar & makeScrollBar(int x, int y, String label) {
        return P.ScrollBar.new(x, y, label);
    }
    AbstractWindow & makeWindow(int x, int y, int h, int w) {
        return P.Window.new(x, y, h, w);
    }
};

WidgetFactory.init(Motif);
\end{verbatim}
\end{quote}

\subsection{Uses}

The Builder pattern~\cite{Gamma_et_al._1995} can benefit from
metaclasses objects in a similar way as the Abstract Factory pattern.
Instead of writing product-specific concrete builder classes, we only
need to parameterize the \verb|Builder| class by the concrete product to be
built.

Interfaces for metaclass objects not only allow us to abstract over
object creation but over any method of the metaclass object.  Suppose,
we want to maintain instantiation counts for several unrelated
classes.  We would define an interface for the methods to access the
instantiation count.  A function to print or analyze these
instantiation counts could then take as an argument any metaclass
object conforming to the interface.
		% Vince
% still need to add references!

% Chambers, Bobrow (CLOS), Michiel & Gabriel (ECOOP 87)
% Steele 90 (CLOS packages)

% check Castagna (covariance <-> multimethods v. two methods)
% dynamic overload resolution?  LFP 92

%packages only?
%inheritance?
%difference to overloading?

%Gerald's comments to Chambers/Leavens

%I don't like the style of having all methods outside the class.  I
%think, it's definitely more readable to have both single dispatched
%methods in a class and multimethods in modules as we propose it.  We
%probably need their algorithm, though, to ensure that multimethods are
%defined for all combinations of argument types.

%To allow programs to be written in the style

%	- define the class hierarchy,
%	- add multimethods to it (Visitor pattern),
%	- extend the class hierarchy,
%	- extend the multimethods

%we might also need their notion of resolving modules and the associated
%link-time test.

\section{Method Dispatching on Multiple Parameters}

\label{section:multimethods}

Typical object-oriented programming languages such as Smalltalk-80~\cite{Goldberg_Robson_1983} and \CPP~\cite{Ellis_Stroustrup_1990} use
{\em single dispatching\/} to determine the method invoked.  When a
method is invoked on a receiver object, a suitable method
implementation is selected dynamically according to the class of the
receiver.  Other arguments do not influence method selection and are
simply passed to the method.

This approach works well for many kinds of methods, especially when
the receiver is more important than the other arguments for
determining which method implementation to select.  However, for some
kinds of methods, there might be several equally important arguments,
and the asymmetry of the single-dispatch style is not appropriate.
Furthermore, it is often cumbersome to add a new method to each class
in an otherwise stable class hierarchy.  Dispatch based on multiple
arguments would allow defining the new method without modifying the
class hierarchy by dispatching on an argument whose type is a class
from this hierarchy.

Methods whose selection is based on several arguments are called {\em
multimethods\/} as supported in CLOS~\cite{Bobrow_et_al._1988, Steele_1990, Paepcke_1993} and Cecil~\cite{Chambers_1992, Chambers_1993}.  Typical multimethods include
binary arithmetic operations, binary equality testing, and
simultaneous iteration over several collections.

\subsection{Examples}

Multimethods are not directly supported by single-dispatching
object-oriented languages, but can be simulated by invoking several
methods such that each argument that participates in method selection
acts as the receiver once.  As soon as the class of an object is
known, the class name is encoded in the name of the next method
invoked.  This technique is called {\em multiple dispatching\/}~\cite{Ingalls_1986} or, in the common case of dispatching on two
arguments, {\em double dispatching}.  Double dispatching is
exemplified in the Visitor pattern~\cite{Gamma_et_al._1995}.

\subsubsection{The Visitor Pattern}

%As the authors say themselves, all we need for that is multiple
%dispatch.  The visitors could then be bundled together in packages.

In the Visitor pattern, a visitor represents an operation to be
performed on the elements of a structure.  This allows defining new
operations without changing the classes of the elements to be operated
on.

As a typical \CPP\ example of the Visitor pattern, consider abstract
syntax trees and operations on these trees in the context of a
compiler (adopted from~\cite{Gamma_et_al._1995}).  Abstract syntax
trees are built from nodes for assignments, variable references,
expressions, and so on.  Operations on abstract syntax trees include
type checking, code generation, flow analysis, etc.

If these operations were defined as methods for each node class, the
code implementing the operations would be distributed over the node
class hierarchy, making it hard to understand and maintain.
Furthermore, adding a new operation would be tedious and would usually
require recompiling all node classes.

Instead, the Visitor pattern expects the node classes to have a single
method called \verb|Accept| that takes a visitor object as an argument.
The pattern further relies on separate visitor classes corresponding
to the operations.  For each node class $X$|Node|, each visitor has a
visitation method \verb|Visit|$X$ that is invoked from within the \verb|Accept|
method in the node class.  This makes it easy to add a new operation,
such as a new code optimization scheme, in the form of a new visitor
class.

When the \verb|Accept| method is invoked, method selection is first based
on the node class; the name of the node class is then encoded in the
visitation method invoked on the visitor object; finally, the
visitation method is selected based on the actual visitor class.
Hence the Visitor pattern provides the effect of double dispatching.

First, we define the \verb|Node| interface with various implementation
classes.  
%
% I'm not sure we really need the Node interface.  It isn't used anywhere.
\begin{quote}
\begin{verbatim}
interface Node {
    void Accept(NodeVisitor & v);
    // ...
};

class AssignmentNode : implements Node {
public:
    void Accept(NodeVisitor & v) { v.VisitAssignment(this); }
    // ...
};

class VariableRefNode : implements Node {
public:
    void Accept(NodeVisitor & v) { v.VisitVariableRef(this); }
    // ...
};
\end{verbatim}
\end{quote}

We define the \verb|NodeVisitor| interface with implementations
|\verb|TypeCheckingVisitor| and {\tt Code\-Generating\-Visitor}.
\begin{quote}
\begin{verbatim}
interface NodeVisitor {
    void VisitAssignment(Node & n);
    void VisitVariableRef(Node & n);
    // ...
};

class TypeCheckingVisitor : implements NodeVisitor {
public:
    void VisitAssignment(Node & n);
    void VisitVariableRef(Node & n);
    // ...
};

class CodeGeneratingVisitor : implements NodeVisitor {
public:
    void VisitAssignment(Node & n);
    void VisitVariableRef(Node & n);
    // ...
};
\end{verbatim}
\end{quote}

Several problems arise when using the Visitor pattern.  First, it is
difficult to add new element classes.  Each new element class
$X$|Node| requires defining a method \verb|Visit|$X$ in the interface
|\verb|NodeVisitor| and corresponding method implementations in each visitor
class to be added.  This makes the visitor class library difficult to
maintain when not only operations, but also element classes are added
frequently.

% ??? do we ever add many element classes ??? only if the abstract
% syntax changes, right?

%4. par (after example): You argue that you want to frequently add element
%   classes.  In that case, you wouldn't use the visitor pattern.  It would
%   be better to argue that you want to BOTH frequently add element classes
%   and frequently add visitors.

Second, the Visitor pattern assumes that the public interface of the
|\verb|Node| class is large enough so that visitors can do their job.  This
often leads to larger interfaces than otherwise desirable and
potentially defeats encapsulation.

Third, double dispatching is error-prone because it requires method
selection to be implemented manually.  This makes it hard to see which
argument combination causes the execution of which method.

%An unrelated problem is that we would like to prohibit certain
%combinations of parameter classes.  For example, we do not want
%ordinary drivers to register for trucks.  Since the method \verb|Accept|
%has to be in the \verb|Node| interface, it would be necessary to raise an
%exception at run-time.

\subsection{Solution}

Simulated double or multiple dispatching is a standard paradigmatic
idiom in languages that do not support multimethods directly.  The
Visitor pattern is actually an instance of this idiom.  

The problems of the multiple dispatching idiom can be avoided if the
language provides multimethods as a built-in construct.  Such a
construct allows the programmer to indicate on which arguments method
selection should be based.  Instead of specifying method lookup
procedurally (as in the idiom), multimethods provide a way of
specifying method lookup declaratively~\cite{Chambers_1992}.

We argue that multimethods make it easy to define new visitation
methods or extend the element and visitor hierarchies alike.
Using \CPP-like syntax, multimethods can be defined as methods in a
class or object.  We suggest to perform dynamic dispatching on all
arguments whose type is a class reference or class pointer.  Unlike
static overloading resolution, dispatching is performed at run-time,
although multimethods can be statically typed~\cite{Rouaix_1990, Castagna_Ghelli_Longo_1992, Chambers_Leavens_1995}.

By combining multimethods and packages, we can simplify the abstract
syntax tree example by bundling the visitation methods for each tree
operation in a package.
\begin{quote}
\begin{verbatim}
object TypeChecker {
public:
    void Visit(AssignmentNode & n) {
        // ...
        Visit (n.LHS ());
        Visit (n.RHS ());
        // ...
    }

    void Visit(VariableRefNode & n);
    // ...
};

object CodeGenerator {
public:
    void Visit(AssignmentNode & n);
    void Visit(VariableRefNode & n);
    // ...
};

Node & root = // ...

TypeChecker.Visit(root);
CodeGenerator.Visit(root);
\end{verbatim}
\end{quote}
Both node classes and visitor packages can now be added easily.  Adding
a node class usually requires adding the corresponding \verb|Visit| methods
in the visitor packages; methods can also be added to existing visitor
packages by using inheritance.  Adding a visitor package does not require
any change to the node classes; however, the new visitor package must
provide a \verb|Visit| method for each existing node class.

Alternatively, the multimethods for type checker and code generator
could have been put into a single package by dispatching on two
arguments.  While not useful for this example, having a single package
makes sense in other cases, for example, when binary methods are
involved.

%The question arises why we still need separate visitor objects.  These
%classes are useful if we need to maintain a state in the visitor
%objects or provide additional functionality beyond the visitation
%function.  In cases where we want to apply only a single visitation
%function to a node, we could pass a function as the second parameter of
%the \verb|Visit| multimethod instead of a visitor object.  In a language
%with closures, the state could be captured implicitly by the closure
%(see also section \ref{section:closures}).

% I prefer to use `parameter' intead of `formal argument'.  Instead
% of `if the actual argument type is an interface' you could say
% `if an interface reference is passed'.  This should get the
% terminology consistent with the rest of the paper and avoid
% any confusion.

The proposed multimethod mechanism follows the general rule that every
method invocation must have a receiver.  For a multimethod defined in
a class, the receiver is an instance of the class.  For a multimethod
defined in an object, the receiver is the object itself.
Syntactically, multimethods appear like statically overloaded methods
in \CPP~\cite{Ellis_Stroustrup_1990}.  Semantically, multimethods
reduce to overloaded methods only when passed a class pointer or
reference.

We propose an {\em exact-match\/} multimethod selection scheme that
operates in two steps.  First, the method is dispatched with respect
to the receiver; this step determines in which class or object the
method is defined.  Second, from all methods within the class of the
receiver or within the receiver object, a method is selected based on
an exact match on the classes of all arguments.  This approach allows
the uniform treatment of all methods as multimethods.

To avoid the complexity of best-match algorithms for multimethod
selection, we suggest dispatching only on parameters that have class
reference or class pointer types.  If the type of an argument in such
a parameter position is an {\em interface\/} reference or pointer
type, then the {\em run-time\/} dispatch mechanism narrows down the
set of suitable method implementations to methods whose parameter type
in this position matches the current class of the argument value.  If
the type of an argument is a {\em class\/} reference or pointer type,
then the set of suitable method implementations can be narrowed down
at {\em compile time\/} to methods with this type in the given
parameter position.  This process is repeated for each argument until
finally a single method is selected to be called.

If all values passed to a method have class types, the method
implementation can be fully selected at compile time.  In this case,
method selection reduces to static overloading.  To avoid the need for
complicated disambiguating rules for which method to select,
parameters that have interface reference or pointer types should not
be used for dynamic method selection, and all multimethods with the
same name should have the same type in this parameter position.  Since
method dispatch is only performed for parameters of a class type,
there can be only one method with a given name whose parameters are
all of interface types.

To guarantee that all method invocations can be handled, it is
necessary to check statically whether all combinations of parameter
classes and interfaces are provided by a multimethod.  Since the
proposed object system achieves subtyping via interface conformance
instead of inheritance, an explicit subtype hierarchy is no longer
available at compile time (see Sections~\ref{section:interfaces}
and~\ref{sec:inheritance}).  Therefore, determining statically whether
a multimethod is defined for all parameter type combinations is
slightly more complicated than in systems with an explicit subtype
hierarchy such as Cecil~\cite{Chambers_Leavens_1995} and would require
linker support.  The necessary information could be collected by the
linker from the compiler-generated method dispatch tables.

%On the other hand, exact-match multimethod selection with linker
%support is capable of detecting undefined combinations of argument
%classes at link-time.  By contrast, best-match selection schemes
%usually resort to a default case in which an exception is raised (see
%also~\cite{Lea_1993}).

%, as illustrated in the example
%given earlier for the Visitor pattern.  In our example, attempts to
%register an ordinary driver for a truck are now detected instead of
%resorting to a default method that raises an exception as in the , as
%illustrated in the same way as in the Visitor pattern, for example, by
%raising an exception (see also
%\cite{Lea_1993}).

To maintain an object-oriented or data-abstraction-oriented view of
multimethods, we could conceptually consider a multimethod as part of
each class for which the method dispatches.  It would thus be tempting
to give each multimethod access to the non-public fields of the
parameters on which dispatching is performed.  However, this would
compromise encapsulation since anyone could now gain access to the
implementation of an existing class by writing a multimethod that
dispatches on a parameter of that class.  We propose separating
dispatching and access as a solution to this problem.  To grant the
multimethod access to non-public fields, the class should declare the
multimethod or an entire package containing the multimethod as its
|\verb|friend|.  Encapsulation problems as encountered in the Visitor
pattern thus no longer occur.

\subsection{Uses}

Typical uses of multimethods include binary arithmetic operations,
binary equality testing, simultaneous iteration over several
collections, and displaying a shape on an output device (see also~\cite{Chambers_1992}).

For example, a multimethod for iterating simultaneously over two
collections could be defined as follows.  This method would traverse
the two collections in lock step and apply a visitation function to
each pair of items visited in each step.  Here, we present only the
case for iterating over two lists.
\begin{quote}
\begin{verbatim}
void PairDo(List & c1, List & c2, PairVisitor v) {
    for (ListIterator i(c1), j(c2); ! i.IsDone() && ! j.IsDone; i.Next(), j.Next())
        v.Apply(c1.CurrentItem(), c2.CurrentItem());
}
\end{verbatim}
\end{quote}

%%% Local Variables: 
%%% mode: latex
%%% TeX-master: "Paper"
%%% End: 
		% Konstantin
\section{Implementation of the Proposed Object Model}

\label{section:implementation}

To illustrate that the proposed set of language constructs and mechanisms fit
together and can be implemented efficiently, this section gives an
overview of techniques for their implementation and presents implementation
issues that arise when combining these constructs and mechanisms.

\subsection{Overview of the Implementation}

Although the proposed forms of inheritance and import do not preclude
a traditional dispatch table implementation of inheritance, they can
be implemented more efficiently and simpler than \CPP-style
inheritance.  Since these forms of inheritance do not allow class
instances to be cast up and down the inheritance hierarchy, and since
non-virtual methods cannot be overridden, a compiler can simply copy
the code of inherited classes and recompile it.  The advantages of
such an implementation are fivefold.  First, all dispatches on {\tt this}
within a method can be statically bound and possibly inline expanded.
Second, virtual method dispatch tables
are no longer needed.  Third, object migration in a distributed environment
is easier to implement since all the code that needs to be shipped is
readily available without walking the class hierarchy.  Forth, generating
code for {\em path expressions\/}~\cite{Campbell_Habermann_1974} or
similar mechanisms for synchronizing concurrent method calls becomes
simpler since a method might have different synchronization
constraints depending on the class of the receiver.  Finally, the
semantics are cleaner than with the complicated table lookup in \CPP.

An implementation of actually copying and recompiling source code has
three disadvantages.  First, it is expensive at compile time since
inherited methods are repeatedly compiled.  Second, it introduces
additional space overhead since the multiple compiled versions of each
method are almost identical.  Finally, it prohibits inheriting classes
for which only header files and object files are provided.

By copying intermediate code (e.g., byte code) or object code,
however, we still have all the advantages compared to an
implementation using dispatch tables without the overhead of
recompilation and the necessity of having source code available.  An
advantage of copying source code or intermediate code rather than
object code is that it enables some compiler optimizations such as
eliminating unused methods and fields or more aggressive inlining.

For type-checking inheritance in the presence of \verb|selftype|, the
compiler has to employ the matching algorithm as described in~\cite{Bruce_1996,Abadi_Cardelli_1995}.

The implementation of interfaces with structural conformance is
straightforward and has been discussed in the literature~\cite{Baumgartner_Russo_1995a,Baumgartner_Russo_1996}.
For an interface declaration, the compiler does not generate any code
but only enters the interface type into the symbol table.  For an
interface pointer/reference initialization or assignment, the compiler
verifies that the RHS type conforms to the LHS type and, if necessary,
generates a dispatch table.  For each method in the interface, the
dispatch table contains the address of the corresponding class method.
If inheritance is not implemented by code folding but rather through
code sharing and a table lookup (as in \CPP), the interface dispatch
table might contain pointers to pieces of code, called {\em thunks}.
These thunks then adjust the object reference as necessary and either
branches to the class method directly or performs a virtual method
call.  An interface pointer or reference is represented as a pair of
pointers, one pointing to the class object, the other pointing to the
dispatch table.

Classless objects do not require any special treatment.  Like a class
instance, a classless object is simply allocated on the heap, or in
static memory if the object is in global scope, and initialized.  A
closure object, like a function closure, has to contain a {\em static
link}, i.e., a pointer to the enclosing function or method scope~\cite{Aho_Sethi_Ullman_1985}.  The compilation of metaclass objects is no
different than that of other objects.

Dispatching a multimethod with $n$ arguments requires an
$n$-dimensional table lookup.  For multimethods in a class, this
lookup could be combined with the dispatch on the receiver of the
method.  However, since this would require all classes participating
in the dispatch to be known, the resulting dispatch table could not
be generated until link time.  Because of this and since such tables
could become very large, it is preferable to keep the multimethod
dispatch separate from the dispatch on the receiver at the cost of an
additional memory reference.  Further techniques for reducing table
sizes are discussed in~\cite{Amiel_Gruber_Simon_1994}.

\subsection{Run-Time Type Identification}

Since inheritance does not define a subtype relationship, an interface
table uniquely identifies an interface-class pair, or an
interface-object pair in case of a classless object.  It is,
therefore, possible to use interface tables for {\em run-time type
identification\/} purposes, i.e., for identifying the implementation
type of an object.

Run-time type identification can be implemented efficiently by
including in the interface table a pointer to the run-time
representation of the implementation type.  In case of a class
instance, the run-time type representation is the class object.
For a classless object, the pointer would point to the object itself.

With run-time type identification, it is possible to implement a
\verb|typecase| statement of the form
\begin{quote}
\begin{verbatim}
typecase (p) {
case A  : /* p points to an A */;  break;
case B  : /* p points to a B */;   break;
case C  : /* p points to a C */;   break;
default : // ...
}
\end{verbatim}
\end{quote}
where \verb|p| is a signature pointer or an expression that evaluates to a
signature pointer and \verb|A|, \verb|B|, and \verb|C| are classes or classless
objects.  If the compiler temporarily enters the implementation type
in the symbol table entry of an interface pointer, a \verb|typecase|
statement can be used to cast an interface pointer to a class pointer
in a type-safe manner.

% RTTI would also allow a polymorphic clone operation in the class
% object.  This can be done by translating
%
%	S * q = clone p;
%
% to
%
%	S * q = p.optr->rtti->clone (*p);
%
% I this useful?  Or should clone be a method of the object?

% There doesn't seem to be any advantage of clone being a method of
% the class object.  In Smalltalk, copy is a method of the object.

Another advantage of run-time type identification is that it
facilitates the implementation of multimethod dispatch.  The pointer
to a class object or a classless object stored in an interface
table can be used as a type identifier.  Depending on the
implementation of multimethod dispatch table, this pointer can either
be used directly as index into the table or it can be used to look up
a type identifier in the class object.

\subsection{Multimethods in the Presence of Structural Subtyping}

With our definition of multimethod selection, a dynamic dispatch
is only performed when an interface pointer is passed to a
multimethod.  For example, assume \verb|C| and \verb|D| are classes conforming
to interface \verb|S| and the multimethod \verb|f| is defined as follows:
\begin{quote}
\begin{verbatim}
int f (C *) { return 1; }
int f (D *) { return 2; }
\end{verbatim}
\end{quote}

For compiling the call
\begin{quote}
\begin{verbatim}
S * p = new C;
// ...
int i = f (p);
\end{verbatim}
\end{quote}
the compiler has to perform a multimethod dispatch to select one of
the two implementations of~|f|.  However, because of structural
subtyping, the compiler cannot guarantee that in the call of~|f| the
interface pointer does not point to an object of any type other than
\verb|C| or \verb|D|\@.

In a language where the subtype relationship is defined by inheritance
or where interface conformance has to be declared by name, the subtype
hierarchy is explicit in the source code.  It is, therefore, possible
for the compiler to test whether a multimethod is defined for all
possible argument combinations.  For example, {\sc Cecil}~\cite{Chambers_Leavens_1995} allows the definitions of a multimethod~|g|
to be spread across multiple modules.  To resolve any possible
inconsistencies between independent modules, the programmer has to
write a {\em resolving module\/} that extends all modules containing
definitions of~|g| and adds any missing definitions.  While the linker
has to verify the existence of the resolving module, the compiler can
can verify the completeness of the set of definitions of~|g| when
compiling the resolving module.

With structural subtyping, the subtype hierarchy is not explicit in
the program, which makes it impossible for the compiler to verify that
a multimethod is defined for all possible combinations of arguments.
In addition, since we require an exact match for selecting a
multimethod to keep the semantics simple, it is not possible to write
a default method.  For example, adding the definition
\begin{quote}
\begin{verbatim}
int f (S *) { return 0; }       // illegal
\end{verbatim}
\end{quote}
to the multimethod~|f| above is not allowed.

The solution is to use linker support for reconstructing the type
hierarchy.  Since the compiler generates a dispatch table for each
pair of types that are in a subtype relationship, the linker can
reconstruct the subtype hierarchy by collecting all interface dispatch
tables.  The linker can then either check the consistency of
multimethods directly or provide the subtype graph to the compiler.

\section{Conclusion}

We have examined design patterns, as representatives of common
programming practice, and analyzed the influence the choice of
implementation language has on pattern design.  We believe this
analysis benefits both the patterns community and the language
communities.

\subsection{Impact on Design Patterns}

Based on our analysis, we have presented several {\em general-purpose\/}
language constructs and mechanisms that, if available in an object-oriented
programming language, would simplify object-oriented design patterns.
We argue that even if they are not available in a chosen
implementation language, design patterns should be described in terms of
these constructs and mechanisms. 

To use design patterns written in terms of a larger set of constructs
and mechanisms than the implementation language supports, it is
necessary to combine these patterns with {\em paradigmatic idioms},
which are patterns implementing the missing language constructs and
mechanisms for a particular implementation language.
This combined approach makes patterns simpler, more succinct, and applicable
to a larger set of language communities.

With our set of language constructs and mechanisms, it is clearly
possible to express all existing patterns since these constructs and
mechanisms are a superset of those found in typical object-oriented
languages.  Additional constructs or mechanisms might be beneficial
but would likely be specific to individual patterns and of less
general-purpose value.

\subsection{Impact on Language Design}

We suggest that future object-oriented languages provide the language
{\em constructs\/} interface, object, and class and the {\em
mechanisms\/} interface conformance, code reuse, lexical scoping, and
multimethod dispatch.

The proposed language mechanisms are all orthogonal; none can be
simulated by a combination of the others.  These mechanisms extend
traditional object-oriented mechanisms.  Multimethod dispatch is an
extension of single dispatch.  We propose separating the subtyping and
code-reuse aspects of inheritance and, as a result, could strengthen
both.

Of the proposed language constructs, interface and object are
orthogonal.  A class is not orthogonal to objects since it also
creates the metaclass object.  Rather a class is syntactic sugar for a
combination of the metaclass object together with a representation
type.  For orthogonality, it might be desirable to add a language
construct for specifying a representation type directly.  Such a
construct would be similar to \verb|record| in Pascal or \verb|struct| in C,
except that it can also contain methods.  However, since the
combination of a representation type together with a metaclass object
containing at least the method \verb|new| is that common, the \verb|class|
construct subsumes the functionality of a \verb|record|.  Since the class
syntax is terser than manually defining a metaclass object together
with a representation type, and since it is more conventional, it
should be included with any object-oriented language.

It is up to the language designers to integrate the proposed
constructs and mechanisms in an orthogonal way into the design of an
actual language.  To our knowledge, there does not yet exist a
programming language that supports the full range of our language
constructs and mechanisms.

Modern functional and abstract data type languages, such as
ML~\cite{Milner_Tofte_Harper_1990,Milner_Tofte_1991},
{\sc Haskell}~\cite{Hudak_Fasel_1992,Hudak_et_al._1992}, and
{\sc Axiom} and its predecessor {\sc Scratchpad~II}~\cite{Jenks_Sutor_1992,Watt_et_al._1987,Sutor_Jenks_1987}, have
closures and packages but lack classes.  {\sc Objective CAML} is an
ML dialect with classes, but it requires explicit coercion instead of
allowing implicit subtyping and it does not offer class interfaces,
metaclass objects, or multimethods.

Most contemporary statically typed object-oriented languages do not
have closures and have only a rudimentary form of packages.
For example, \CPP~\cite{Koenig_1995,Ellis_Stroustrup_1990} and {\sc Modula-3}~\cite{Cardelli_et_al._1992} lack class interfaces with structural
subtyping, metaclass objects, and multimethods.  {\sc Java}'s object
model~\cite{Sun_Microsystems_1995} is similar to that of \CPP\ and
adds interfaces without structural subtyping.  {\sc Emerald}~\cite{Black_et_al._1986,Black_et_al._1987} has interfaces, objects,
and structural subtyping but does not have classes, inheritance, or
multimethods.  POOL-I~\cite{America_van_der_Linden_1990} has
interfaces separate from classes and structural subtyping but does not
have classless objects, metaclass objects, or multimethod dispatch.
{\sc Cecil} separates interfaces from implementation and subtyping
from inheritance and provides packages, closures, and multimethods but
does not offer classes in the traditional object-oriented sense, class
inheritance, singleton objects, or metaclass objects.

A language based on our constructs and mechanisms would benefit a
larger community than just the object-oriented community.  For
example, using lexically scoped closure objects allows us to write
programs in an abstract data type style or in a functional style as
well as in an object-oriented style.

\bibliographystyle{plain}
\bibliography{patterns,signatures}

\end{document}